\documentstyle[12pt]{article}

\large

\def\be{\begin{equation}} \def\ee{\end{equation}}
\def\ba{\begin{eqnarray}} \def\ea{\end{eqnarray}}

\def\ts#1{\textstyle{#1}}
\def\bra#1{\langle#1\vert}\def\ket#1{\vert#1\rangle}
\def\ev#1{\langle #1 \rangle}\def\tE{\tilde E}
\def\dd{{\rm d}}\def\e{{\rm e}}\def\DD{{\scriptscriptstyle D}}
\def\half{{\textstyle{1\over2}}}\def\ovr{\overline}

\begin{document}

\title{Statistical Geometry of Random Weave States}

\author{Luca Bombelli\\
Department of Physics and Astronomy,\\
University of Mississippi, University, MS 38677, USA}
\date{{\small 15 January 2001}}

\maketitle

\begin{abstract}
\noindent I describe the first steps in the construction of semiclassical
states for non-perturbative canonical quantum gravity using ideas from
classical, Riemannian statistical geometry and results from quantum geometry
of spin network states. In particular, I concentrate on how those techniques
are applied to the construction of random spin networks, and the calculation
of their contribution to areas and volumes.
\end{abstract}

\subsection*{Introduction}

\noindent The non-perturbative canonical approach to quantum gravity has made
a great deal of progress over the past decade, with a few of the main
developments being a rigorous construction of the kinematical Hilbert space
in the connection representation, results on the quantum geometry of states
in this Hilbert space, and the proposal of a Hamiltonian operator for the
theory \cite{[Horowitz]}. Although the quantum theory is not yet complete
even in the pure gravity case, enough of it is in place for us to start
developing rigorously the semiclassical theory and its coupling to matter,
with the aim of addressing its phenomenological aspects, some of which have
already been treated in the literature \cite{[Phenom]}, but only in a
heuristic way.

A basic ingredient for such developments is an understanding of semiclassical
gravitational states and their geometry. At the heuristic level, these
considerations led several years ago to the proposal of weave states
\cite{[Weaves]}. Starting from loop states \cite{[Loops]}, which encode
distributional information about the spatial geometry along the
one-dimensional submanifolds defined by a set of (possibly intersecting)
loops, weave states were based on uniformly distributed loops in the spatial
manifold, and sought to encode in some sense all of the geometry down to the
length scale associated to the loop spacing.

Analogs of the weave states in the context of the rigorous formulation of the
quantum theory are natural candidates for semiclassical states, and some
papers have recently appeared looking at this question \cite{[Semicl]}. Among
the first questions about such states that need to be addressed are their
definition and relation to the kinematical Hilbert space, and the nature and
extent of the geometrical information they encode. These issues are being
studied in a collaboration with A. Ashtekar \cite{[StatGeom]}, and I will
describe here some of the work, touching on the more geometrical aspects of
our proposal, and using as the main example a flat geometry.

The basic, canonically conjugate variables of the theory are a
density-weighted triad field $\tE^a_i$ on the spatial manifold $\Sigma$ (the
``electric field''), from which the inverse of the spatial metric $q_{ab}$
can be obtained as $(\det q)\,q^{ab} = \tE^a_i \tE^b_j\, \delta_{ij}$, and an
su(2)-valued connection $A_a^i = \Gamma_a^i + k\, K_a^i$, where $\Gamma_a^i$
is the connection with respect to which $\tE^a_i$ is covariantly constant,
$K_a^i$ corresponds to the extrinsic curvature $K_{ab}$ of a constant $t$
slice on a classical solution, in the sense that $K_a^i = (\det q)^{-1/2}
K_{ab}\,\tE^b_i$, and $k$ is an arbitrary number, the Immirzi parameter.

In the connection representation, pure states in the kinematical Hilbert
space $\tilde{\cal H}^0$ of the theory (i.e., before imposing the quantum
constraints) are given by functionals $\Psi[A]$ of suitably generalized
connections $A$ on $\Sigma$, and a basis for $\tilde{\cal H}^0$ is given by
the spin network states \cite{[Spinnets]}, defined by triples $S = \{\gamma,
{\bf j}, {\bf I}\}$, where $\gamma$ is a directed graph embedded in $\Sigma$,
${\bf j} = \{j_e\}$ a set of labels for representations of SU(2), one for
each edge $e$ of $\gamma$, and ${\bf I} = \{I_v\}$ a set of labels for
intertwiners at all vertices $v$, i.e., a (gauge-invariant) prescription for
contracting the indices of the matrices $R^{(j_e)}$ that the representations
$j_e$ associate with the holonomies $U(e,A)$ of $A$ along the edges $e$
incident at each $v$. For every such triple, a wave functional in the
connection representation for quantum gravity is defined by
\be
   \Psi_{\gamma,{\bf j},{\bf I}}[A]
   = \langle\gamma,{\bf j},{\bf I}\mid A\rangle
   := \prod\nolimits_e\prod\nolimits_v I_v R^{(j_e)}\big(U(e,A)\big)\;.
   \label{sn}
\ee
On these states, of course, $\hat A_a^i$ is a multiplication operator,
$\hat E^a_i$ a differential one.

In terms of (\ref{sn}), weave states have a graph $\gamma$ given by a
uniformly distributed collection of loops, with respect to a classical
geometry $(\Sigma,q_{ab})$, and all holonomies are taken in the
$j=\half$ representation. For these states, it was found that they were
eigenstates of the operators $\hat A_S$ and $\hat V_R$ corresponding to areas
of surfaces $S$ and volumes of regions $R$ in $\Sigma$, as is true more
generally for spin network states, and that the eigenvalues would be the
correct areas and volumes given by the classical metric, for large surfaces
and regions, if the loops were uniformly scattered with mean spacing
$a = \sqrt{2\pi}\, \ell_{\rm P}$ between them, with $\ell_{\rm P}$ the Planck
length, which was seen as a first indication of a fundamental discreteness in
the non-perturbative quantum theory, emerging from the theory itself as
opposed to being put in by hand.

In general, in order to ask whether a quantum state $\Psi$ is semiclassical,
a set of observables $\{g_\alpha\}$ must have been previously specified; the
state is semiclassical if the expectation values of the corresponding set of
operators coincide with the values assigned by those observables to a
classical phase space point $(\tE,A)$ on $\Sigma$, $\ev{\hat g_\alpha}_\Psi =
g_\alpha(\tE,A)$, and the uncertainties are small, in the sense that $(\Delta
g_\alpha)_\Psi\ll g_\alpha(\tE,A)$, or $\ev{\hat g_\alpha^2}_\Psi \ll
[g_\alpha(\tE,A)]^2$ (except when the classical value itself is small). A
proposal for a set of semiclassical states will then start with the choice
of a sufficiently large, and physically relevant, set $\{g_\alpha\}$.

In the weave state approach, the choice was $\{g_\alpha\} = \{A_S,V_R\}$
for large and ``slowly\ varying'' $S$ and $R$, and we will make the same
choice here. Other choices are possible, and one is illustrated by the
coherent state construction proposed by T. Thiemann and collaborators
\cite{[Coherent]}, in which the basic observables are the holonomies of $A$
along the edges of a given graph $\gamma$, and what can be viewed as fluxes
of the electric field through a given set of surfaces, each one intersecting
transversally one of the edges of the graph.

In this paper, we give a procedure for constructing random, uniformly
distributed graphs in $\Sigma$ that can be used to define random spin
networks. Although the procedure is more generally applicable, we treat the
case of a flat metric, and use results from classical statistical geometry of
Euclidean manifolds and from quantum geometry of spin network states to find
the contribution to areas and volumes from states of this type; such states
are eigenstates of those geometrical operators, and can be thought of as
corresponding to the previous, heuristic weave states. We then comment on the
possibility of using those states to construct more general ones, and on the
extension to curved geometries.

\subsection*{Classical Statistical Geometry: Random Complexes}

\noindent Consider a Euclidean manifold $(\Sigma, e_{ab})$, of finite volume
$V_\Sigma$ and arbitrary dimension $D$, for the time being. We start by
sprinkling $N$ points at random in $(\Sigma, e_{ab})$, independently and with
uniform density $\rho = N/V_\Sigma$; that is, the probability density that
each sprinkled point fall at any location $x$ in $\Sigma$ is
\be
   \dd P(x) = {\dd v\over V_\Sigma}
   = {\ts{\sqrt{e(x)}}\over V_\Sigma}\,\dd^\DD x\;, \label{dP}
\ee
and the probability that it fall in any (measurable) region $R \subset \Sigma$
is therefore $V_R/V_\Sigma$ (all volumes $V_R = V(R)$ will be defined using
$\dd v = \sqrt e\,\dd^\DD x$ from now on.) Combining these probabilities for
single points we find that, when the $N$ points have been chosen, the
probability that exactly $n$ of them be in any given $R\subset\Sigma$ is
given by the distribution
\be
   P_{\rm binomial}(n,R|N,\Sigma)
   = {N\choose n}\left({V_R\over V_\Sigma}\right)^{\!n}
   \left(1-{V_R\over V_\Sigma}\right)^{N-n}. \label{Pn}
\ee
To perform an actual sprinkling in a computer simuation is very easy if the
geometry is flat, since then coordinate values can be chosen uniformly at
random in a Cartesian chart; if the geometry is curved, other well-known
techniques can be applied (see, e.g., Ref.\ \cite{[Recipes]}).

Two special limiting cases deserve mention. One is the continuum limit,
approached as $N$ and $\rho$ become very large, with constant $V_\Sigma$;
the other is the infinite volume limit, in which $N$ and $V_\Sigma$ are
very large, with $\rho = N/V_\Sigma$ a constant. Let us analyze the latter
situation in more detail. Since $V_\Sigma = \infty$, in this case we cannot
use the probability density (\ref{dP}); we do, however, have probabilities for
finite regions. It is a standard, well-known result, that in the infinite
volume limit the distribution (\ref{Pn}) approaches a Poisson distribution
with mean $\ovr n = \rho\,V_R$,
\be
   P_{\rm binomial}(n,R|N,\Sigma)
   \approx P_{\rm Poisson}(n|\rho V_R)
   = {\e^{-\rho V_R}\big(\rho\,V_R\big)^n\over n!}\,;
   \label{Poisson}
\ee
for this reason, uniform random distributions are often called Poisson
random lattices \cite{[Fields]}. Thus, simulating a random sprinkling in a
region $R$ which is part of an infinite-volume manifold $\Sigma$ is a two-step
process, in which one first generates the number $N_R$ of points in $R$ using
the distribution $P_{\rm Poisson}(N_R|\rho V_R)$, and then generates
locations for those points inside $R$ as in the compact manifold case. Of
course, the same two-step procedure can be followed when a compact manifold
needs to be divided into two or more disjoint regions for the simulation,
although in that case the binomial distribution is used for the first step.

Regarding fluctuations in $n$, the standard deviation of a Poisson
distribution is well-known, $\sigma_n^{\rm Poisson} = \sqrt{\ovr n} =
\sqrt{\rho V_R}$; for the binomial distribution, one can readily verify that
\be
   \sigma_n^{\rm binomial} = \sqrt{\rho V_R\left(1-{\rho V_R\over N}\right)}
   \approx \sigma_n^{\rm Poisson} \left(1-{\rho V_R\over2N}\right),
\ee
where in the last, large $N$ approximation we have neglected terms of order
$N^{-2}$. Similar results hold for individual probabilities, which justifies
the use of the Poisson distribution as an approximation even when the number
$N$ of points in a compact manifold is fixed; one may say, e.g., that the
probability that the region $R$ contain no sprinkled points is $\e^{-\rho
V_R}$, using Eq.\ (\ref{Poisson}), provided $N$ is large.

Once an arbitrary, locally finite (assumed here to be random) distribution of
points $\{p_i\}$ is given in a Euclidean manifold $(\Sigma, e_{ab})$, there
is a well-known construction, often called Dirichlet-Voronoi construction, of
two cell complexes based on that distribution: a simplicial complex $\Delta$
triangulating $\Sigma$, with those points for vertices, and its dual cell
complex $\Omega$.

The simplicial complex $\Delta$ can be obtained in the following way. We
associate a simplex with a subset of $D+1$ points among all the $\{p_i\}$
if the (unique) S$^{\DD-1}$ sphere passing through all of them does not
contain any other point $p_j$ (this sphere is defined by the metric $e_{ab}$,
and has as its center the unique point which is equidistant from the $D+1$
$p_i$'s). The set of all such simplices is the desired complex; it covers
$\Sigma$, and any two of them can only have a vertex, an edge, or a 2-face in
common \cite{[Lee]}. Being made of simplices, this complex has a fixed number
of $l$-faces in each $k$-face, for $l<k$ (for example, in $D=3$ dimensions,
there are exactly 4 triangles, $l=2$, in each tetrahedron, $k=3$), but the
number of $l$-faces sharing a given $k$-face, for $l>k$, depends in most
cases on the specific set of points used, and for us will be a random
variable (like, e.g., again for $D=3$, the number of edges, $l=1$, sharing a
given vertex, $k=0$).

The cell complex $\Omega$, dual to $\Delta$, is obtained by defining, for
each point $p_i$, a cell $\Omega_i$ to be the set of all manifold points
which are closer to $p_i$ than to any other $p_j$; the set of all such cells
as $p_i$ varies is the desired complex \cite{[Lee]}. Therefore, each $D$-cell
$\Omega_i$ of this complex is dual to a vertex $p_i$ of the simplicial
complex, and each $(D-1)$-cell $\Omega_{ij}$ is dual to an edge $p_{ij}$, and
perpendicular to it if they meet.\footnote{With the above, standard
construction of $\Delta$ and $\Omega$, not all edges meet their dual faces.
One possible slight modification of the construction is to use, given a set
of sprinkled points, the same simplicial complex $\Delta$, but choose as
vertices of the dual complex the incenters (centers of the inscribed spheres)
of all simplices, as opposed to the circumcenters (centers of the
circumscribed spheres), which is what the standard procedure amounts to. The
topology of the resulting dual complex $\Omega'$ is the same as that of
$\Omega$, but in the pair $(\Delta,\Omega')$ dual elements always intersect,
although in general not perpendicularly.} In general, there is a duality
between $k$-faces $p_{i_1...i_k}$ in the simplicial complex and
$(D-k+1)$-cells $\Omega_{i_1...i_k}$, and the incidence relations reflect this
duality. Thus, there is a fixed number of $l$-cells sharing each $k$-cell,
for $l>k$ (for example, in $D=3$ dimensions, there are exactly 4 edges
sharing each vertex, except for degenerate cases), but the number of
$l$-cells in each $k$-cell, for $l<k$, in most cases depends on the specific
set of points used, and for us will be a random variable (like, e.g., again
for $D=3$, the number of faces in each 3-cell). We denote the set of
$k$-faces of $\Delta $ by $\Delta^{(k)}$, and the set of $l$-cells in
$\Omega$ by $\Omega^{(l)}$.

When the above constructions are used with uniform random distributions of
points of density $\rho$, one gets random simplicial and dual cell complexes
$\Delta_\rho$ and $\Omega_\rho$, that have been studied for a long time. In
the context of gauge theory, their use was proposed in the early 80's as a
way of implementing a short distance cutoff without breaking Euclidean
invariance \cite{[Christ],[Lee]}, but results on statistical properties of
random complexes had been obtained long before in metallurgy and mineralogy
\cite{[Minerals]}, motivated by studies of crystal formation by random
nucleation in minerals, and by mathematicians \cite{[Santalo]}.

For our applications to quantum gravitational states, two kinds of related
properties of the above complexes will be important. Incidence relations
between simplices or cells of different dimensionalities, that are
statistical {\it topological\/} properties of the complexes, will relate the
number of $D$-simplices to the number of sprinkled points, and therefore give
us the density of dual cell complex vertices in terms of $\rho$. On the other
hand, {\it metric\/} properties, such as average cell sizes, will be more
closely related to average intersection numbers of cells of various
dimensionalities with given subsets of $\Sigma$, such as the number of
intersections per unit area between a given surface and graph edges. All
properties of this type are known for two- and three-dimensional Euclidean
space; we will discuss here those properties we will need later for our
quantum statistical geometry results.

To calculate incidence relations, we consider the (finite) simplicial complex
$\Delta_\rho$. Denote by $N_k$ the total number of $k$-faces of $\Delta_\rho$;
in particular, $N_0 = N$ is the number of points. Two relations among the
$N_k$'s can be used in any dimension. One is the expression for the Euler
characteristic of a simplicial complex, $\sum_{k=0}^\DD(-1)^kN_k =
\chi(\Delta_\rho)$, where the dependence on the global topology of $\Sigma$
can be eliminated by dividing the equation by $N$, and then taking the large
$N$ limit, in which $\chi(\Delta_\rho)/N\to0$, and
\be
   1-{N_1\over N_0}+{N_2\over N_0}-\cdots+(-1)^\DD{N_\DD\over N_0} = 0\;.
\ee
The other relation is the fact that each $D$-simplex has $D+1$ faces of
codimension 1, each one shared by two $D$-simplexes, so that
\be
   N_{\DD-1} = {D+1\over2}\,N_\DD\;.
\ee
In $D=2$ dimensions, these two equations would be sufficient for determining
the average number of edges and 2-simplices per unit volume, which in the
uniform sprinkling case are the mean densities of the corresponding
simplices, constant throughout $\Sigma$; in $D=3$ dimensions, however, we
need more equations. A detailed calculation, where one integrates
explicitly over the probabilities of finding points at various locations
(see, e.g., Ref.\ \cite{[Fields]}), shows that each cell has on average
$96\pi^2/35$ vertices; since each of those vertices is shared by four cells,
and there are $\rho$ cells per unit volume, we find the density of cell
complex vertices to be
\be
   {N_3\over V_\Sigma} = {1\over4}\,{96\pi^2\over35}\,\rho
   = {24\pi^2\over35}\,\rho\;. \label{Nvert}
\ee

To calculate the metric properties of random complexes, we must make
extensive use of the probabilities associated to the point distribution. In
$D=3$ dimensions, we are interested in the mean number of edges intersected
by a flat surface per unit area, for which we can use the following argument.
As has long been known (see, e.g., references in \cite{[Minerals]}), from a
generalization of the Buffon needle method of calculating $\pi$ by random
tosses of a stick on a series of parallel lines, when a randomly oriented set
of lines of arbitrary (possibly disconnected) shape and total length $L$,
contained in a region of volume $V$, is cut by a surface of area $A$, the
number of intersections between the lines and the surface per unit area is
very simply related to the line length per unit volume, $\ovr{N_{\rm int}/A}
= \half\,L/V$. The problem of finding the mean number of intersected edges is
thus reduced to that of finding the mean total edge length per unit volume. A
known calculation gives that the mean edge length per 3-cell is $(4\pi)^{5/3}
(3^{1/3}/5)\,\Gamma({4\over3})\,\rho^{-1/3}$, from which $\bar L/V$ can be
found multiplying by the mean number of cells per unit volume, $\rho$, and
dividing by the number of cells sharing each edge, 3. Putting all of this
together then gives
\be
   \ovr{N_{\rm int}/A}
   = {(4\pi)^{5/3}\,3^{1/3}\over30}\,\Gamma(\ts{4\over3})\,\rho^{2/3}
   \approx 2.917\,\rho^{2/3}\;. \label{Nint}
\ee
This result does not depend on the shape of the surface (in particular,
$S$ does not have to be flat), but other moments of the probability
distribution for $N_{\rm int}$, such as its width, do.

\subsection*{Quantum Statistical Geometry: Areas and Volumes}

\noindent In this section, we start discussing how to use the random cell
complexes introduced above to construct quantum gravity states. The amount of
structure we use from each pair of complexes is a choice we make in each
approach, depending largely on the set $\{g_\alpha\}$ of observables we want
to reproduce. Since our observables here are areas and volumes, results from
quantum geometry that we will summarize below allow us to make the simplest
possible choice: We use only the 1-skeleton, or set $\gamma_{\bf x} =
\Omega^{(1)}_\rho$ of edges of the dual cell complex obtained with a
randomly generated set of points ${\bf x} = \{x_i\}$, as a random spin network
graph. The main advantage of this choice, as opposed to using the simplicial
complex $\Delta_\rho$, is the fact that almost every vertex is exactly
$(D+1)$-valent in $D$ dimensions; notice that, given that the edges are
geodesic segments, the degree of differentiability of the graph depends on
the geometry, and in the present case the graph is piecewise analytic.

In this paper, we will also make other simplifying choices, namely those of
using only one such graph at a time for the states we consider, and of
defining a spin network state $\Psi_{\gamma,{\bf j},{\bf I}}$ with the same
spin label $j$ at each edge and the same intertwiner $I$ at each vertex; we
can then write our states as $\Psi_{\gamma,j,I}$. These further choices are
made purely for illustrative purposes here, and will be changed in a more
complete treatment; a few comments on this point will be included in the final
section, and see Ref.\ \cite{[StatGeom]} for details. This amounts to assuming
that any single random graph $\gamma$ provides a good sampling of the
underlying manifold, and we end up with a two-parameter family of states for
each such graph. One question we can address is then what constraints are
placed on the parameters by the requirement that the area and volume
operators have the right values on these states.

Consider a state $\Psi_{\gamma,j,I}$ of the type just described. Given a
surface $S$ in $\Sigma$, any given vertex of the random graph $\gamma$ will
fall on $S$ with probability zero, and we can consider all intersections of
$S$ and $\gamma$ to be single, transversal edges. In this case the spin
network is an eigenstate of the area operator with eigenvalue
\cite{[Area]}
\be
   A_S = 8\pi k\,\ell_{\rm P}^2
   \sum\nolimits_\alpha\sqrt{j_\alpha(j_\alpha+1)}\;, \label{As}
\ee
where $j_\alpha$ is the half-integer label for the $\alpha$-th edge crossing
$S$. If all the $j_a$'s are equal to a given $j$, the area eigenvalue for a
given spin network becomes $8\pi\,k\,\ell_{\rm P}^2\,N_{\rm int}
\sqrt{j(j+1)}$, where $N_{\rm int}$ is the number of intersections between $S$
and the graph, so from Eq.\ (\ref{Nint}) we obtain that on average classical
and quantum areas agree if $\rho$ and $j$ satisfy
\be
   {(4\pi)^{5/3}\,3^{1/3}\over30}\,\Gamma(\ts{4\over3})\,\rho^{2/3}
   = \left(8\pi k\,\sqrt{j(j+1)}\,\ell_{\rm P}^2\right)^{-1}\;.
   \label{areas}
\ee

Given a region $R\subset\Sigma$, spin network states are eigenvectors of the
volume operator $\hat V_R$, and the corresponding eigenvalues receive a
contribution from each vertex of the graph,
\be
   \hat V_R\Psi_{\gamma,j,I} = \kappa_0\,(8\pi k)^{3/2}\,\ell_{\rm P}^3
   \sum\nolimits_v\sqrt{|\hat q_v|}\,\Psi_{\gamma,j,I}\;,
\ee
where the constant $\kappa_0$ is an undetermined factor arising from a
regularization ambiguity for the volume operator, and $\hat q_v$ is an
operator corresponding to the determinant of the spatial metric at $v$, whose
eigenvalues are determined by the $j_\alpha$'s of all edges incident at that
vertex and the intertwiner $I_v$ \cite{[Volume]}. While a closed formula
like the one for area eigenvalues, Eq.\ (\ref{As}), is not available for
volumes, the calculations simplify in the case of four-valent vertices. When
the four $j$'s are equal, the number of independent intertwiners at $v$, and
thus of eigenvalues of $\hat q_v$, is $j+1$; when $j$ is even, one of those
eigenvalues is zero, while the others, and all eigenvalues when $j$ is odd,
come in pairs of opposite signs and therefore give doubly degenerate volume
eigenvalues. The volume eigenvalues have been calculated for the first few
values of $j$ \cite{[Loll]}; for example, the only eigenvalue for $j=1$,
and the non-zero ones for $j=2$, are
\be
   \lambda_1 = {3^{1/4}\over4}\,\kappa_0\, k^{3/2}\,\ell_{\rm P}^3\;,
   \qquad
   \lambda_2 = {3^{1/4}\over2}\,\kappa_0\, k^{3/2}\,\ell_{\rm P}^3\;.
\ee
In general, we can write the eigenvalues for vertices of our type in the form
\be
   \lambda_{j,I} = f(j,I)\,\kappa_0\, k^{3/2}\,\ell_{\rm P}^3\;,
\ee
so from Eq.\ (\ref{Nvert}) we obtain that on average classical and quantum
volumes agree if
\be
   {24\pi^2\over35}\,\rho
   = \left(f(j,I)\,\kappa_0\, k^{3/2}\,\ell_{\rm P}^3\right)^{-1}\;.
   \label{volumes}
\ee
If the parameters $k$ and $\kappa_0$ of the theory are known, equations
(\ref{areas}) and (\ref{volumes}) are constraints on the parameters
characterizing the random spin network states. Notice, however, that more
constraints on the latter parameters are expected to arise from matching the
manifold curvature in more general cases, in which we consider non-flat
manifolds $(\Sigma,q_{ab})$, and we can also think of those two equations as
fixing $\kappa_0$ once values for $j$ and $I$ have been determined,
\be
   \left(\sqrt{j(j+1)}\,{(4\pi)^{8/3}\,3^{1/3}\over15}\,
   \Gamma(\ts{4\over3})\right)^{3/2} = {24\pi^2\over35}\,f(j,I)\,\kappa_0\;.
\ee

\subsection*{Concluding Remarks}

\noindent An obvious way to generalize the states used in this paper is to
relax the assumption that all $j$'s and all $I$'s on a given graph $\gamma$
be equal, and define a state
\be
   \ket{\gamma,C} = \sum\nolimits_{{\bf j},{\bf I}}
   C_{{\bf j},{\bf I}}\,\ket{\gamma,{\bf j},{\bf I}}\;,
\ee
for some set of coefficients $C_{{\bf j},{\bf I}}$ to be determined. But the
more interesting modification to our states comes from the observation that,
given $(\Sigma, e_{ab})$, the sprinkling process does not give a unique set of
points in $\Sigma$, but rather a probability density for $N$-point
distributions,
\be
   \dd P_\rho(x_1,...,x_N)
   = N!\prod_{i=1}^N{\sqrt{e(x_i)}\over V_\Sigma}\,\dd^\DD x_i\;,
\ee
parametrized by the density $\rho$, or $N=\rho V_\Sigma$, and therefore the
dual cell complex construction also gives us a probability density on the set
of graphs embedded in $\Sigma$ depending on the parameter $\rho$, rather than
a single graph. We can then use this probability density to integrate over
random graphs, and obtain either pure states resulting from their
superposition, or mixed states of the form
\be
   \Psi_e(\rho,C) = \int\dd P_\rho(x_1,...,x_N)
   \,\ket{\gamma_{\bf x},C}\bra{\gamma_{\bf x},C}\;, \label{mixed}
\ee
where the subscript $e$ makes the dependence on the metric $e_{ab}$ explicit.
Superposing graphs in this way may seem like a complication, but on the other
hand it has the advantage that, while a single $\ket{\gamma_{\bf x}, {\bf j},
{\bf I}}$ would only have an approximate Euclidean invariance, the integral
is exactly invariant (despite having an ultraviolet cutoff scale), as well as
covariant with respect to the action of diffeomorphisms.

In the previous section, we have only discussed (for the simpler states
$\Psi_{\gamma,j,I}$) the expectation value of areas and volumes, and seen how
they give rise to conditions on the parameters the states depend on. The
uncertainties that states of the form (\ref{mixed}) associate with those
observables, or others we may use to identify semiclassical states, will have
a quantum contribution and a classical, statistical one; imposing that they
be smaller than the desired tolerance will introduce further conditions, and
restrictions on the length scales defined by the geometry and the subsets of
$\Sigma$ we consider.

In addition, it is useful to keep in mind alternatives to some other choices
we made; although they were the simplest ones, there is no guarantee that once
we understand the dynamical aspects better they will appear as the best ones.
Specifically, in a curved manifold, uniform distributions of points are not
the only possible covariantly defined ones; a uniform distribution in a
Riemannian manifold $(\Sigma, q_{ab})$ is equivalent to using the density
$\dd v = \sqrt{q}\, \dd^\DD x$ as a measure $\dd\mu$ on $\Sigma$, while
alternative ones can be defined using any scalar constructed from the metric,
such as the Ricci scalar, $\dd\mu = R\, \dd v$, or any other curvature
scalar. Possibly related to this is the fact that a better understanding of
the theory may show that, in order for a spin network state to have a fully
consistent semiclassical interpretation (including, e.g., the fact of giving
rise to distributions for the spatial geometry and its time derivative which
are both peaked around classical values), the weave we use must be
constructed using correlations in the sprinkling process, that were not used
here.

\subsection*{Acknowledgements}

\noindent In addition to Abhay Ashtekar, I would like to thank in particular
Jurek Lewandowski and Thomas Thiemann for a series of stimulating discussions
and suggestions.


\end{document}